# Particle beam experiments for the investigation of plasma-surface interactions: application to magnetron sputtering and polymer treatment

Carles Corbella,[a)] Simon Grosse-Kreul, Oliver Kreiter, Teresa de los Arcos, Jan Benedikt, and Achim von Keudell

*RD Plasmas with Complex Interactions, Ruhr-Universität Bochum, Universitätsstr. 150, 44780, Bochum, Germany*

A beam experiment is presented to study heterogeneous reactions relevant to plasma-surface interactions. Atom and ion beams are focused onto the sample to expose it to quantified beams of oxygen, nitrogen, hydrogen, noble gas ions and metal vapor. The heterogeneous surface processes are monitored in-situ and in real time by means of a quartz crystal microbalance (QCM) and Fourier transform infrared spectroscopy (FTIR). Two examples illustrate the capabilities of the particle beam setup: oxidation and nitriding of aluminum as a model of target poisoning during reactive magnetron sputtering, and plasma treatment of polymers (PET, PP).

---

[a)] Author to whom correspondence should be addressed. Electronic mail: carles.corbella@ruhr-uni-bochum.de.

# I. INTRODUCTION

Glow discharges at low pressure are the nucleus of many industrial applications, mainly thin film growth, plasma etching and surface modification.[1] Knowledge of elementary processes taking place at surfaces in contact with plasmas are of crucial importance to tailor the physical and chemical properties of materials. Ion bombardment and chemical reactions on the surface determine to a great extent the mechanical properties of hard coatings,[2,3] the extent of polymer cross-linking and roughness during plasma treatment,[4,5] and the oxidation processes undergone by targets during reactive sputtering.[6] The use of quantified particle beams permit to analyze the heterogeneous processes occurring during plasma-surface interactions experimentally.[7,8] Different technical approaches and facilities specifically dedicated to vacuum beam experiments can be found in the literature:

- The experiments of Winters and Coburn pioneered the interaction between etchants (F, Cl, H and combinations) and silicon, and they served as motivation for the first complete beam setup for plasma research.[9] Plasma etching processes were simulated by exposing samples to beams of Ar ions, and to molecules and atoms of different reactive elements. Surface diagnostics were performed by quartz microbalance (QCM), Auger Electron Spectroscopy (AES) and two Quadrupole Mass Spectrometers (QMS), one of them with several stages of differential pumping. Several samples could be mounted on a carousel and were bombarded by different particle beams. Dual atom beams were produced by two Evenson cavities and could be modulated with a chopper wheel. An ion gun generated $Ar^+$ beams at some hundreds of eV.[10] The basic scheme of beam experiment has bee transferred to different reactors. Kimura et al improved the existing setup by incorporating a neutralizing filament and a transfer rod to introduce the samples in the beam chamber.[8]

- A variant of Atomic Layer Deposition (ALD) is constituted by Radical Enhanced ALD (REALD), which has been studied by beam experiments in the specific case of TiN film growth.[11] Independent control of the beams of $TiCl_4$ molecules, deuterium and nitrogen radicals has been achieved, so that TiN is deposited by the repetition of three sequential steps of beam bombardment. TiCl has been used as liquid precursor and was delivered with a doser. Film growth was also performed onto a QCM cooled down with liquid nitrogen flowing inside the rotary carousel (Dewar).

- The measurements of absolute densities of radicals and molecules have been the leading motivation in molecular beam epitaxy growth. In the work of Chen et al, vacuum-UV (VUV) absorption spectroscopy and quadrupole mass spectrometry were combined to study high density radical sources.[12]

- Interaction of Materials with Particles and Components Testing (IMPACT) is a sophisticated system equipped with many in-situ techniques, like X-ray Photoelectron Spectroscopy (XPS), Auger Electron Spectroscopy (AES) and mass spectrometry.[13] Ion and electron spectroscopies are performed using a hemispherical energy multichannel analyzer. A dual control unit of quartz crystal microbalance (DCU-QCM) monitors the mass variation rates. Among the various particle sources, high- and low-energy ion sources; metal ion sources by either electron impact ionization of metal vapor or by thermionic process; X-ray source, and an extreme UV (EUV) source can be identified. An electron-beam evaporation source is also available to deposit ultrathin multilayer films.

- An antecedent of IMPACT is provided by MAJESTIX. Chemical sputtering and growth of carbon coatings is the major research line of this instrument.[7,14] Hydrogenated or deuterated carbon films were deposited by radiofrequency (RF) plasma-enhanced chemical vapor deposition (PECVD) in the preparation chamber, which acted also as load-lock chamber. Real-time diagnostics by spectroscopic ellipsometry and FTIR in reflection mode were carried out using two lines of sight. The samples were exposed to two radical beam sources (45º) and one ion gun (perpendicular incidence). The column of the ion source was equipped with a Wien filter and deceleration optics, which provided ions with energies down to 1 eV. Methyl radicals are obtained by thermal dissociation of azomethane.

- Plasma Etching and Deposition ReactOr (PEDRO) is a versatile beam and plasma system devoted to the fabrication of metal contacts and hard coatings on ion treated substrates. Its flexible configuration of three moving electrodes facing the center of a spherical main chamber permits to combine magnetron sputtering, PECVD and ion bombardment.[15,16] Plasma and ion beam characterizations are performed with a retarding field analyzer and a Faraday cup.

- PISCES-B is aimed to investigate fusion plasma interactions with materials. Energy control of the impinging ions is performed by scanning the substrate bias. Concretely, the irradiation of beryllium with deuterium is one of the most significant breakthroughs.[17] Carbon erosion experiments have been also conducted in PISCES-B, like in the case of MAJESTIX.[18]

- Another fusion-oriented system is the Garching LArge DIvertor Sample test facility (GLADIS). Equipped with two high-energy hydrogen beam sources, it is conceived to test the thermomechanical behavior of samples undergoing extreme thermal loads.[19] Power densities of up to 50 MW/m$^2$ are supplied to the investigated samples, whose state is monitored by an infrared camera, pyrometers and arrays of thermocouples.

- The basic construction of Magnum-PSI in Rijnhuizen finished in 2009. A very dense magnetized plasma beam in a 15-meter-long setup is generated to study the contact issues between fusion plasmas and a wall. This experiment is conceived to test the response of the wall material at plasma densities and temperatures close to the ITER divertor region at smaller scales.[20] Hot ionized gas are generated by a cascaded-arc plasma source and confined by a very strong magnetic field (higher than 2 T). Diagnostics of plasma state are performed by Thomson scattering and optical emission spectroscopy, whereas target surface is characterized by laser-induced desorption.

Plasma diagnostics at the surface level are routinely performed by substituting the film substrate or target surface by a Faraday cup, quadrupole mass spectrometer (QMS)[21,22] or retarding field energy analyzer (RFEA).[23,24] Aside from the measurement of plasma parameters, in situ monitoring of the chemical state at the sample surface is gaining interest nowadays, since it complements the characterization of the plasma-surface processes. For instance, real-time tracking of the optical properties and the chemical composition and bonding at the surface can be performed by in-situ spectroscopic ellipsometry and Fourier transform infrared spectroscopy (FTIR).[14] In another example, quartz crystal microbalance (QCM) constitutes a widespread tool to measure the deposition rate of thin films.[25] Since QCMs show resolutions in film thickness lower than 1 Å (10 Å = 1 nm), they provide reliable gravimetric measurements of very high sensitivity. Studies on purification treatments of carbon nanotubes and determination of the erosion rate and oxidation state of metals prove the versatility of QCMs as diagnostic instrument in many research areas.[26-28] In short, real time monitoring with QCM and FTIR in combination with particle beams can provide information about relevant surface processes.

This work describes a particle beam setup, whose successful performance has been demonstrated in previous papers.[29-33] Here we consider two important applications in plasma technology, where surface processes were mimicked by vacuum beam experiments: (1) target poisoning in reactive sputtering, and (2) plasma treatment of polymers. The proposed methodology consists in designing particle beam experiments using different targets and the adequate particle sources. On one side, the investigation of Al oxidation and nitriding shows the potential of these experiments to simulate heterogeneous processes with a multibeam system. On the other hand, we evidence the feasibility of real-time measurements by monitoring the chemical state of polymer thin films, like polyethylene terephthalate (PET) and polypropylene (PP), by FTIR during ion bombardment in reactive atmospheres. We will show how beam experiments monitored in-situ by QCM and FTIR are very appropriate to explore elementary mechanisms of sputtering and deposition in real time.

## II. PARTICLE BEAM SETUP

Fig. 1 shows the side and top views of the reactor where the particle beam experiments were performed. The beam chamber consists of a cylindrical vessel of 20 cm in inner diameter, with a load-lock attached to it. Gases are introduced to the beam chamber by mass flow controllers and are pumped downstream by a parallel combination of a turbomolecular pump and an ion-getter-pump equipped with a titanium sublimation pump. The base pressure is <$10^{-6}$ Pa. The vacuum system has been baked out at temperatures between 120 and 150ºC for 8 hours in order to reach ultra-high-vacuum conditions with the operation of the ion-getter-pump. The working pressure is always of the order of a few $10^{-2}$ Pa, and the mean free path of the particles (≈300 mm) is larger than the distances between sources and target (≤ 90 mm). Therefore, the majority of particles reach the target without undergoing collisions with the molecules of the background gas. It is important to note that the impurity level in the beam setup is not governed by the residual contamination in the UHV at base pressure, but rather by the purity of the gases introduced via the beam sources.

Up to four particle sources provide ion and atom beams focused on a grounded target, which consists in either a QCM (QCM holder) or another sample (FTIR holder) both sited on an XYZ manipulator on the axis of the beam chamber. The distances between sources and the center of the beam chamber range between 70 mm and 90 mm. A shutter built in front of the target is used to protect it from the particle beams and to measure the ion current. An electric resistance installed on the FTIR holder permits to heat up the samples. Moreover, the FTIR sample holder is rotary and it can also be tilted to facilitate the alignment of the IR beam. This sample holder is additionally equipped with a Faraday cup (Kimball Physics FC-70 Series), which is used to measure the ion flux density in front of the ion source.

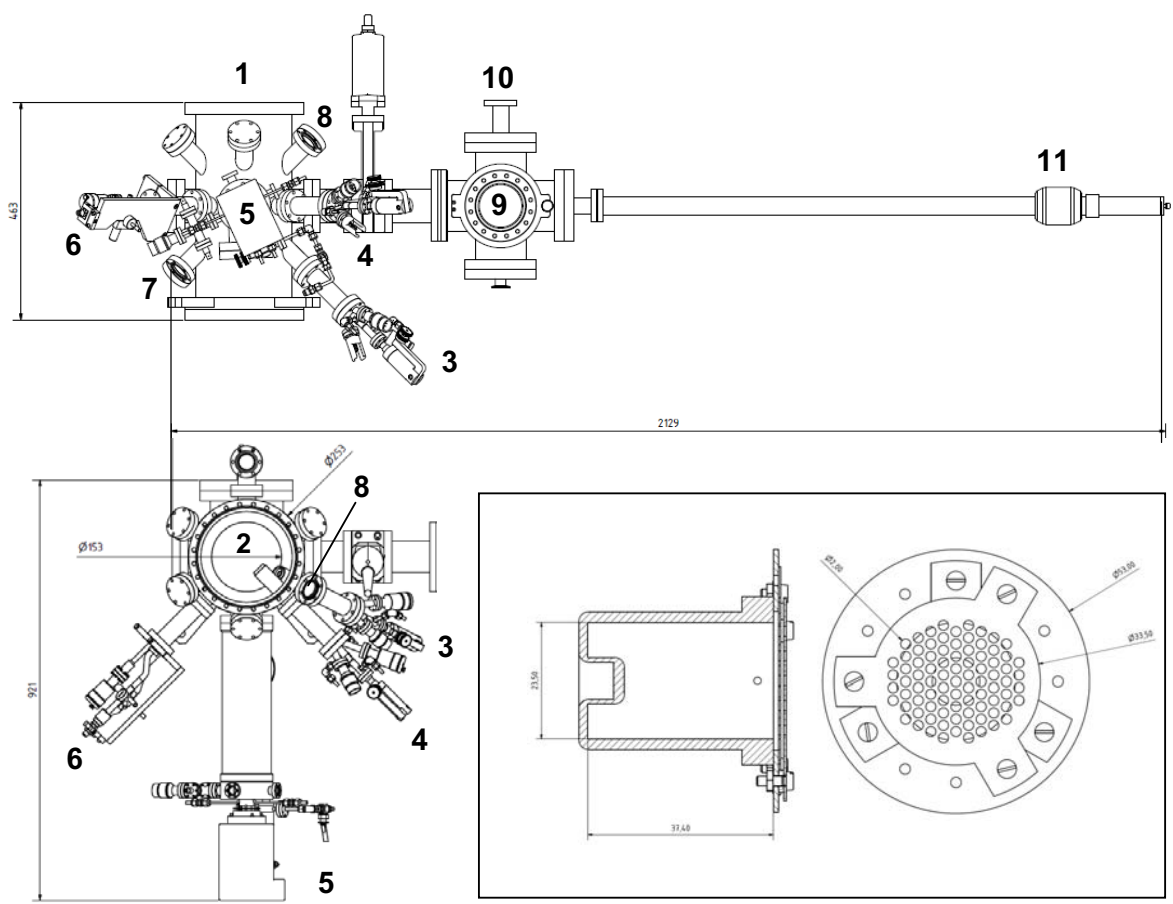

FIG. 1. Side and top view (at scale) of the vacuum beam reactor. List of components: (1) beam chamber; (2) QCM/sample holder; (3) metal vapor source; (4) hot capillary (OBS); (5) ion beam source; (6) Evenson source; (7,8): KBr windows; (9) Load-lock chamber; (10) vacuum carrying case, and (11) transfer rod. Inset: Cross section and front view of the small plasma cup in the ion gun.

The base pressure of the load-lock chamber is ≈$10^{-4}$ Pa. The samples are introduced into the beam chamber by means of a magnetically driven transfer rod. The load-lock admits the incorporation of a vacuum carrying case to transfer the treated samples in vacuum to other systems with compatible ports, such as an X-ray photoelectron spectrometer and an Atomic Force Microscope (AFM), for example. The residual pressure inside the carrying case is around 0.01 Pa within the first 3 hours after being extracted from the load-lock. The conversion of this residual pressure into a gas collision rate yields a monolayer growth time of ≈1 min if we assume a sticking coefficient of 0.01 for contaminants on the surface.

### A. Ion beam source

The plasma ion gun (type Gen 2 by Tectra GmbH) produces ions in an energy interval which can be set between 20 and 2000 eV. 2.45 GHz microwaves are coupled into a small magnetized plasma cup to excite an Electron Cyclotron Resonance (ECR) plasma (inset in Fig. 1). The ions from the plasma cup are extracted using a double-grid system. The bias grid, $V_{bias}$, determines the ion energy, whereas the extraction grid, $V_{ext}$, is used to focus and accelerate the ion beam. The ion flux density lies tipically between $0.5 \cdot 10^{14}$ and $5 \cdot 10^{14}$ cm$^{-2}$ s$^{-1}$ at the sample surface. A circular aperture of 10 mm in diameter was added at the exit of the ion gun in order maintain a plasma pressure of around 1 Pa while keeping the pressure in the beam chamber at 0.02 Pa. The divergence in the ion beam has been characterized with the Faraday cup (Fig. 2a). However, the Faraday cup is placed 4 mm closer to the ion source than the sample and, therefore, the measured ion fluxes are higher. The diameter (FWHM) of the ion beam at the target position is around 20 mm, as determined from the etch profile of a hydrocarbon film deposited onto 26 x 26 mm$^2$ c-Si wafer after being bombarded by Ar ions at 400 eV. The spatial profile shows axial symmetry and the ion flux is uniform along the surface area of the sample where the mass uptake is measured by QCM or the surface chemistry is monitored by infrared spectroscopy.

It is worth noting that the absolute particle fluxes in beam experiments can be lower compared to the plasma processes counterpart that is mimicked. The transfer of any results from the beam experiment to the plasma situation may still be valid, because most heterogeneous surface reactions are dominated by the ion-to-neutral flux ratio only, and typical balance equations to model these processes are linear in the incident particle fluxes.

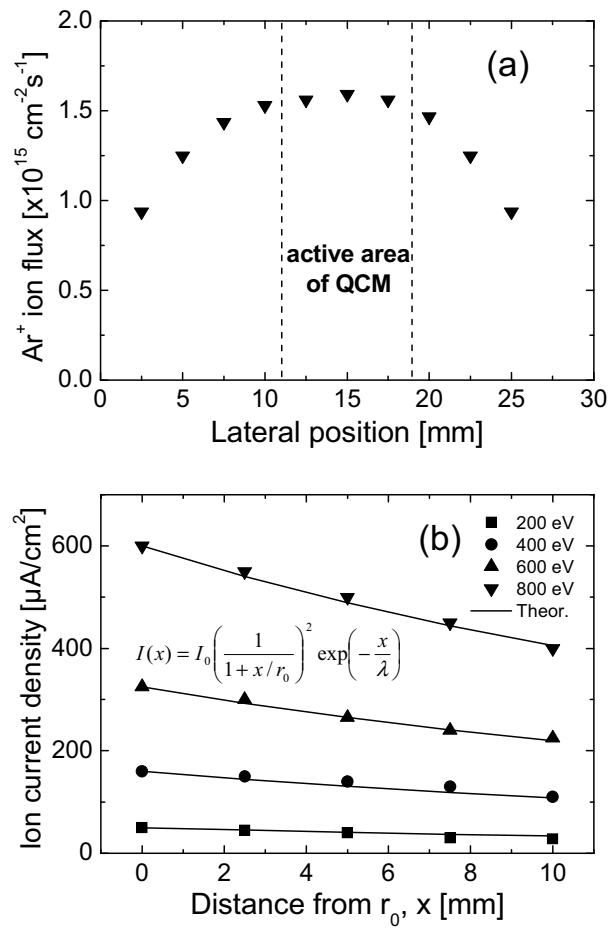

FIG. 2. (a) Spatial profile of the ion flux measured with a Faraday cup (800 eV). (b) Trend of the ion current density at different distances from the ion gun. The constants of the theoretical curves are $x_0$=50 mm and $\lambda$=290 mm.

All the measurements were performed by admitting an Ar gas flow of 1.0 sccm to the plasma ion source. This resulted in a partial pressure of 0.024 Pa in the beam chamber, which is translated into a mean free path of ca. 290 mm.[27] The ion current density was measured at different ion energies by placing a Faraday Cup on the symmetry axis and at different distances from the ion source, as displayed in Fig. 2b. The ion fluxes are compared to a model based on an intensity variation depending on the (1) beam divergence, and (2) collisions with the background gas. The geometric factor of divergence assumes that the ion gun acts as a point source:

$$I(x) = I_0 (1 + x/x_0)^{-2} \exp(-x/\lambda) \tag{1}$$

Here, $I_0$ is the ion current at an initial distance $x_0$=50 mm from the ion gun. $I(x)$ is the ion current at a distance $x$ shifted from $x_0$, and $\lambda$ is the mean free path. The characteristic curves fit well with the experimental data for a given $\lambda$=290 mm.

In a double-gridded ion gun, ions generated in the plasma cup are accelerated within the space between the two grids and exit the gun at a kinetic energy equal to the potential difference set by the grids plus the plasma potential. Ideally, the ions should impinge onto the grounded target at ion energy $e(V_{bias}+V_p) \approx eV_{bias}$, as plotted in Fig. 3a. If charge exchange collisions occur in the acceleration stage, ions at lower energies appear. If we estimate a mean free path of 6.9 mm for Ar ions at 1 Pa, i.e. the approximated pressure within the ion gun,[34] 100 ions undergo 15 collisions in the 1 mm grid separation resulting in 85% of the ion beam flux being unaffected by charge exchange collisions. The remaining ions (15%) undergo collisions contributing to the ion energy distribution (IED) at energies below $eV_{bias}$. Fig. 3b shows the IED of Ar ions at bias voltages of 30 V, 59 V and 108 V. The measurements were performed with a Semion RFEA from Impedans.[23] The plasma potential of the ion source (≈15 V) becomes visible by a shift of IED peak maxima with respect to the adjusted bias voltages, $V_{bias}$. The FWHM ranges between 10 and 15 eV.

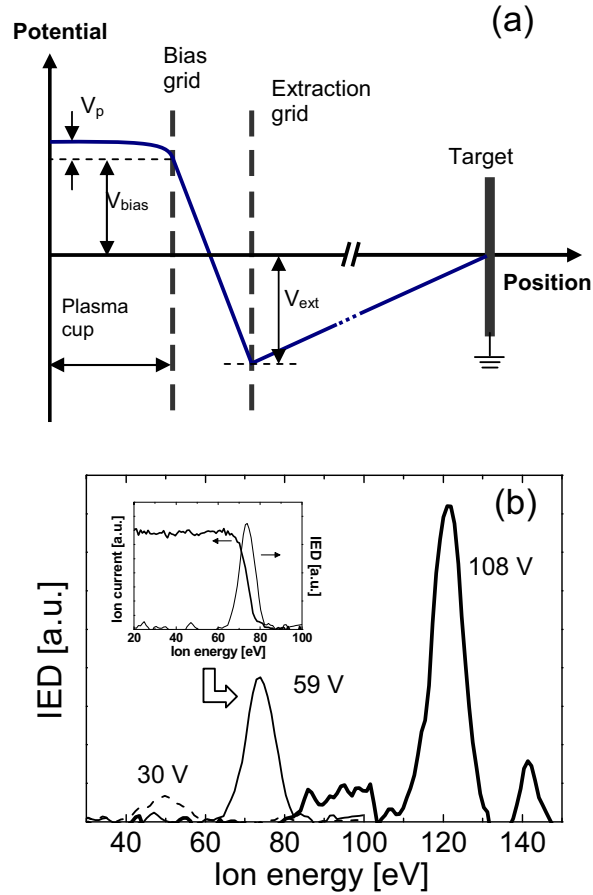

FIG. 3. (a) Schematic representation of the potential evolution between ion gun and target. In general, $V_{bias} >> V_p$. (b) Ion energy distributions of $Ar^+$ beams with bias voltages set at 30 V, 59 V and 108 V. Inset: ion current and the corresponding distribution.

The ECR plasma of Ar generated in the ion gun is also an important source of vacuum-UV (VUV) radiation at wavelengths of 104.9 and 106.7 nm.[1] This VUV background may affect the surface processes that are studied, especially in the case of polymers and resins. The VUV effect can be isolated by suppressing the ion flux emanating from the source by an electrostatic lens in front acting as an ion screener. As an alternative, an ion beam bending section could be installed in front of the plasma source to separate ions from the VUV photons. Such an upgrade of the system is planned in the future.

**B. Hot capillaries: oxygen and hydrogen beam sources**

The Oxygen Beam Source (OBS) (Dr. Eberl MBE-Komponenten GmbH) produces a focused beam composed of atomic and molecular oxygen incident at 45º to the target.[35] This source consists of an iridium capillary which, heated up tol 1800ºC, acts as cracking tube for oxygen molecules flowing inside. At this temperature, the dissociation degree is 15% according to the supplier. Thus, the composition of the oxygen beam is 15% (O) and 85% ($O_2$). In addition, some $O_2$ may impinge from the diffuse background due to the steady partial pressure of $O_2$ in the beam chamber at a typical pressure of a few $10^{-3}$ Pa. This estimate reveals that the oxygen molecule flux from the background is similar to the sum of the direct fluxes of O and $O_2$ from the capillary source.[27], which indicates that the oxygen beams used in the following experiments are composed of a mixture of oxygen molecules and atoms with a flux ratio of $j_{O2}/j_O \approx 10$. This molecular background has to be taken into account for surface systems sensitive to the impact of molecules such as bare metals. In case of semiconductor surfaces, the cross sections for oxidation are much smaller and only the impact of the atoms alone governs the surface reactions. Expressed in absolute fluxes, a molecular oxygen flux ranges between $10^{16}$ and $10^{17}$ $cm^{-2}s^{-1}$, at a flow of 0.1-0.7 sccm $O_2$ through the capillary source. The atom species reach the target with energies corresponding to 1800°C and a pronounced directionality of the beam. A Hydrogen Beam Source (HBS) is also available, whose working principle is equivalent to the OBS but the capillary is made of tungsten.[36]

**C. Evenson cavity: oxygen and nitrogen beam source**

This atom source, which is oriented 45º with respect to the target, is based on a microwave plasma excited in an Evenson cavity.[37] The gas flows through an S-bent quartz tube with a length of 30 cm and 9 mm in inner diameter with a shutter at the end. Molecular dissociation of the reactive gases is promoted within a glow discharge in an Evenson cavity, which is cooled by compressed air. The supplied microwave energy is coupled to the cavity by changing the resonator volume using a manual stub tuner. The quartz tube narrows just after the cavity to increase the pressure inside the plasma region while maintaining a low pressure in the beam chamber. This constriction also reduces the back flow of gas from the beam chamber into the microwave plasma source. Moreover, the S-shaped design of the quartz tube avoids direct exposure of the treated sample to the VUV radiation from the discharge. The dissociation efficiency depends on the supplied power, the gas flow and the injected gas. The fluxes of atomic oxygen at the target generated by the microwave source using gas flows between 3 and 5 sccm are comparable to the ones obtained from the hot capillary. The calibration was performed by measuring the etching rate of a soft hydrocarbon film at a substrate temperature 200°C sequentially exposed to the Evenson source and the OBS, as reported elsewhere.[29]

One advantage of the microwave plasma source is that it is able to dissociate species which cannot easily be cracked by thermal dissociation. The plasma ignited in the microwave cavity is used to produce atomic nitrogen, which is not possible to obtain by thermal dissociation due to the high bonding energy of the $N_2$ molecule. By assuming a maximal dissociation degree of 4% at a supplied power of 40 W,[38] the flux emanating the source ranges between $0.72 \cdot 10^{17}$ and $1.21 \cdot 10^{17}$ cm$^{-2}$s$^{-1}$ at flow interval of 3-5 sccm. An estimation of the collision rate at the target requires information about divergence of the beam. Considering that the dominant contribution of incident species on the target owes to the background gas, the direct flux impinging the target surface can be at least one order of magnitude below than the flux estimated at the outlet of the source.

**D. Metal vapor source**

A beam of metal atoms is generated by an effusion cell WEZ (Dr. Eberl MBE-Komponenten GmbH) consisting in a "cold-lip" pyrolytic BN crucible of 2 cm$^3$. In case of aluminum, the crucible is partially filled with Al pellets (Aldrich Chemistry, 99.999%) as vapor source. The effusion cell is installed under the OBS (Fig. 1), and generates a beam of Al atoms that impinge the QCM with an incidence angle of 70° to the normal of sample surface. The orientation of the crucible has to be carefully selected so that the molten aluminum cannot exit the source as a liquid. A cooling shroud and a shutter are integrated in the effusion cell. Although the melting temperature of Al is 670ºC, the vapor pressure starts increasing significantly from ~900ºC. Fig. 4a shows the aluminum evaporation measured with the QCM. The deposition rate follows approximately an exponential curve at source temperatures between 900°C and 1200ºC. The working temperatures are selected between 1050°C and 1150ºC to provide an aluminum atom flux comparable to the flux of the other incident species in the beam chamber (Fig. 4b).

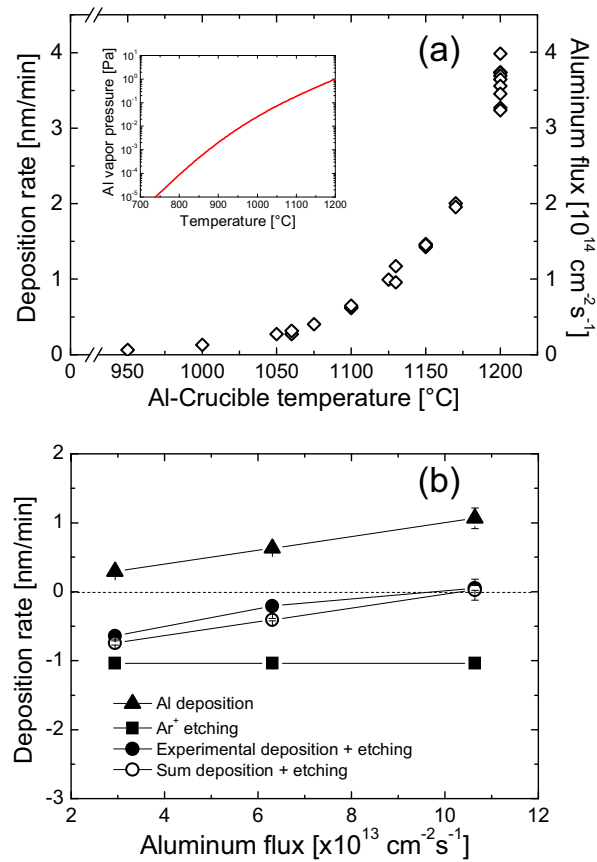

FIG. 4. (a) Calibration curve of the Al-evaporator measured with the QCM. Inset: tabulated evolution of the Al vapor pressure with temperature.[39] (b) Mass variation rates during sputtering of Al by $Ar^+$ etching at 400 eV.

Fig. 4b shows the etching/deposition rates measured with the QCM at different Al-crucible temperatures. First, Al sputtering by $Ar^+$ etching was performed at 400 eV of ion energy. Ar flow was set to 1.0 sccm leading to a pressure of 0.06 Pa in the beam chamber. The argon ion sputter rate is approximately 1 nm/min. Second, the Al deposition rate was measured for the effusion cell operated at 1060 ºC ($2.9 \cdot 10^{13}$ $cm^{-2}s^{-1}$), 1100 ºC ($6.3 \cdot 10^{13}$ $cm^{-2}s^{-1}$) and 1130ºC ($10.6 \cdot 10^{13}$ $cm^{-2}s^{-1}$). The crucible temperatures were regulated according to the calibration curve in Fig. 4a. Finally, both etching and metalizing were carried out at the same time. The sputter rate due to the incident $Ar^+$ ions and the deposition rate due to Al adsorption can simply be summed up to explain the mass variation rate as registered by the QCM during the combined impact.

## III. DIAGNOSTICS

An appropriate in-situ surface diagnostics requires characterization techniques that are non-invasive, that show high resolution and that are reproducible. Surface characterization by QCM and FTIR meet these criteria. On one side, QCM measurements provide mass variation rates during sputtering and deposition processes. On the other hand, FTIR spectroscopy measures the absorption bands on the target during plasma treatment, characterizing the active bonding groups at the sample surface.

### A. Quartz crystal microbalance (QCM)

The thickness changes in the beam setup were measured by a quartz microbalance (QCM). The QCM (Maxtek BDS-250) consists of circular slabs of AT-cut quartz resonators with 0.5 mm thickness and 14 mm diameter. The exposed surface is a circular area with 8 mm diameter. The piezoelectric crystal shows a resonance frequency of 6 MHz. These slabs are coated by metallic layers on each side playing the role of electric contacts. The rear electrode is a gold layer, whereas the front side (exposed to the beams) is coated with an Al film of 1 μm thickness. The QCM measurements are very sensitive to the crystal temperature. Therefore, the QCM holder must be water-cooled during the whole experiments in order to avoid any thermal drifts. The error bars are lower than 1 Hz with a stable temperature, which allows measuring mass variations with submonolayer resolution. Two QCMs are installed in the holder (dual unit), and their resonance frequencies are recorded simultaneously. A built-in two-position shutter selects the crystal to be exposed to the particle beams. The control unit of the QCM consists of an SQC-310 Series Deposition Controller from Inficon.

The frequency shift of the QCM is converted into mass variation, $\Delta m$, by means of the Sauerbrey equation:[25]

$$\Delta m = -N_q \rho_q A \Delta f / f_q^2 \qquad (2)$$

where $N_q$ is the frequency constant of an AT-cut quartz resonator (1.668·10$^5$ Hz cm), $\rho_q$ is the quartz density (2.65 g/cm$^3$), $A$ is the active area (2.01 cm$^2$) and $f_q$ is the resonant frequency without attached mass. The thickness variation rates and the atomic fluxes during deposition/etching are obtained from the mass variation rate, $\Delta m/\Delta t$, by considering the atomic mass and the density of the deposited/etched film.

To test the performance of the QCM upon exposure to the particle beams, mass variation rates of a Diamond-Like Carbon (DLC) thin film (200 nm) were measured by sending an Ar$^+$ beam of 400 eV in presence of O$_2$ injected by the OBS. The DLC films were deposited on the QCM by means of RF PECVD using another reactor in capacitive configuration with acetylene atmosphere at 1 Pa. The Ar gas flow was set to 1.0 sccm all the time. Fig. 5 shows the thickness variation of DLC during Ar/O$_2$ sputtering at different O$_2$ fluxes. Here, a mass density of 2 g/cm$^3$ was assumed.[40] Several stages are marked in the figure, corresponding to variations in the impinging oxygen flux. The first stage (step 1), where only Ar$^+$ ions were used, comprised two regimes differentiated by the slope in the etching curve: cleaning of the passivated surface followed by physical sputtering. The first minutes of etching showed a small variation probably indicating a balance between the mass uptake due to argon implantation and the argon sputtering. The jump at the beginning might be caused by a thermal change due to the effects of both heat of recombination and exposure to the VUV photons emitted by the ECR-plasma of Ar. After 10 min of sputter steady state, we additionally introduced 0.2 sccm of $O_2$ (step 2). The film etching at constant rate was approximately three times faster than in step 1. In the following steps (3 and 4), the oxygen flow was increased to 0.5 and 0.7 sccm. The progressive increase in etching rate in presence of $O_2$ is attributed to the chemical sputtering of carbon due to the formation of volatile $H_2$, CO, $CO_2$ and $H_2O$ molecules.[41] Finally (step 5), the DLC film was etched again only by the $Ar^+$ beam, recovering an etch rate similar to that in step 1 but not identical probably due to residual oxygen in the chamber, aside of oxygen atoms implanted in the DLC film in previous steps. These results have been compared to literature values in Table I.

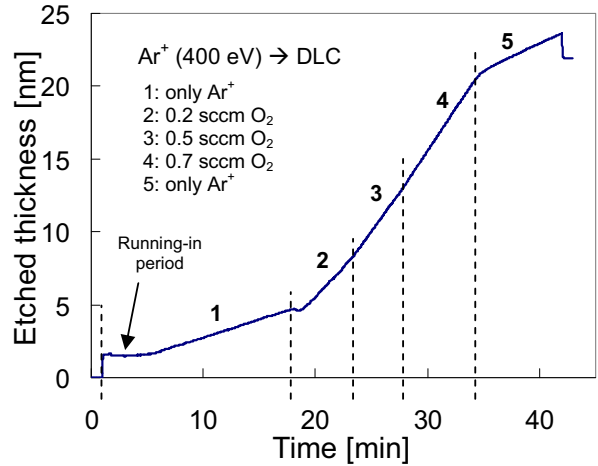

FIG. 5. Etched thickness of the DLC film exposed to an $Ar^+$ beam with 400 eV at different $O_2$ fluxes.

The information displayed in Fig. 5 can be expressed in terms of the effective sputter yield, $Y_{eff}$, which is the relation between the flux of sputtered C atoms, $\Gamma$, and the $Ar^+$ flux density, $j_{Ar}$: $Y_{eff}=\Gamma/j_{Ar}$. $Y_{eff}$. Fig. 5 shows that such yield monotonically increased with the oxygen concentration in the beam chamber. The calculated values of $Y_{eff}$ are listed in Table I, and they have been compared to the results of Hopf et al using the ratio $R=j_{O2}/j_{Ar+}$ as parameter.[42] Our measurements agree with the data of Hopf et al within the $R$ interval 0-36. Moreover, here we provide more accurate sputter yields since the rates have been evaluated in real time. An underestimation of the real sputter yield may be introduced by the ion energy reduction associated to charge exchange collisions in the ion gun.

TABLE I. Study of the chemical sputtering of DLC film depending on flux ratio $R=j_{O2}/j_{Ar+}$. Effective carbon sputter yields, $Y_{eff}$, were calculated at $Ar^+$ flux density, $j_{Ar+}$, $3 \cdot 10^{14}$ $cm^{-2}s^{-1}$, and assuming a hydrogen content ≈30% in DLC. $Y_{eff}$' values at flux ratios $R$' are extracted from Hopf et al.[42]

| Step | $O_2$ flow [sccm] | $O_2$ flux density [$10^{15}$ $cm^{-2}s^{-1}$] | C flux density [$10^{14}$ $cm^{-2}s^{-1}$] | Etching rate [nm/min] | R | $Y_{eff}$ | R' | $Y_{eff}$' |
|---|---|---|---|---|---|---|---|---|
| 1 | 0.0 | 0 | 0.57 | 0.25 | 0 | 0.19 | 0 | 0.28 |
| 2 | 0.2 | 2.68 | 1.9 | 0.82 | 8.94 | 0.62 | 10.1 | 0.75 |
| 3 | 0.5 | 7.51 | 2.4 | 1.06 | 25.0 | 0.81 | 23.8 | 1.05 |
| 4 | 0.7 | 10.7 | 2.7 | 1.16 | 35.8 | 0.88 | 33.6 | 1.17 |
| 5 | 0.0 | 0 | 0.83 | 0.36 | 0 | 0.28 | 0 | 0.28 |

## B. Fourier transform infrared spectroscopy (FTIR)

Etching and chemical modification of the samples are monitored in situ with a FTIR spectrometer (Bruker IFS 66) in reflection mode at an incidence angle of 70º (Fig. 6). The IR beam enters and leaves the main chamber through KBr windows. The beam reflected at the sample is registered by a Mercury Cadmium Telluride (MCT) detector cooled down with liquid nitrogen and located outside the reactor. The IR beam path was continuously purged with compressed air to minimize the IR-absorption by $CO_2$ and water. In some cases, a continuous pumping of the liquid nitrogen cooled MCT detector might be advised to avoid condensation of residual water on the MCT chip inside the detector. Such an ice layer causes a broad absorption peak around 3400 cm$^{-1}$.

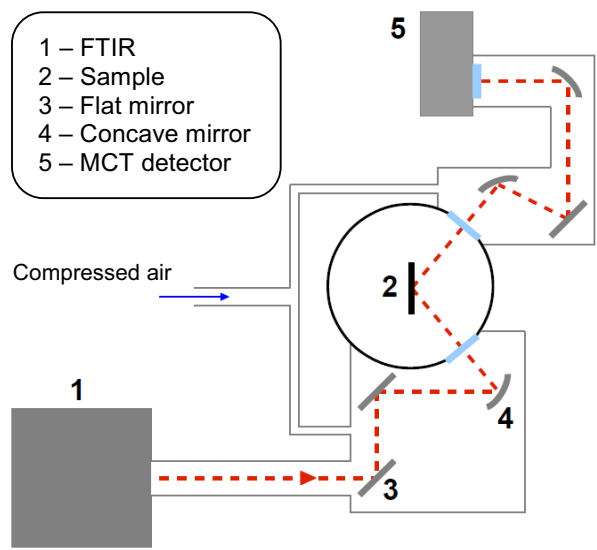

FIG. 6. Sketch of the external IR pathway for the in-situ FTIR measurements.

A very high sensitivity in FTIR measurements is required to analyze the first nanometers corresponding to the interaction depth of the incident species. To enhance the sensitivity in IR absorption, Optical Cavity Substrates (OCS) have been used consisting of an oxidized silicon wafer with an aluminum backside coating. The thickness of the oxide layer is adjusted to a specific thickness to be resonant with IR probing wavelengths of interest. The OCS method can resolve reflectivity changes that correspond to the removal of submonolayers of the material deposited on the top. This configuration increases substantially the IR sensitivity being equivalent to other schemes such as multiple internal reflections (MIR).[43]

The IR reflection spectrum between 400 and 6000 cm$^{-1}$ of a PP spin-coated film, depicted in Fig. 7a, shows roughly a Planck's curve corresponding to black body radiation of the IR emitting source in the FTIR spectrometer. The main absorption peaks are labeled. The noisy modulations between 3500 and 4000 cm$^{-1}$, as well as the peaks in the interval 1500 and 2000 cm$^{-1}$ (1) correspond to stretching vibrations of water. The neighboring bands between 3000 and 3500 cm$^{-1}$ (2) are generated by ice condensed on the MCT detector window due to residual water contamination. The peaks between 2800 and 3000 cm$^{-1}$ (3) correspond to CH$_x$ vibrations from absorbance of protective layers and contaminations on the windows and mirrors in the IR pathway, as well as from the PP probed sample. The strongest absorption in the spectrum at 2350 cm$^{-1}$ (4) owes to CO$_2$ along the IR pathway. Finally, the absorption at around 1030 cm$^{-1}$ (5) corresponds to SiO$_2$ from the OCS. In the case of PET, an additional pronounced absorption peak due to C=O double bond appears embedded within the water absorption region at around 1720 cm$^{-1}$.

Fig. 7b shows a typical series of normalized spectra from a PP spin-coated film being submitted to Ar$^+$ bombardment at 200 eV. Before the treatment, a reference spectrum $R_0$ is measured (Fig. 7a). Then, the FTIR spectra $R$ are continuously sampled during the exposure of the target to the particle beams. 200 scans were averaged every 30 s and they constituted one measurement. The time interval between the spectra displayed in Fig. 7b is 150 s. Each sample spectrum is divided by the reference spectrum so that the etching of the film becomes visible as an increase in reflectivity at specific characteristic IR absorption bands of the polymer. After switching on the Ar$^+$ beam, $R/R_0$ increases at the spectral region of the CH$_x$-band between 2800 and 3000 cm$^{-1}$, indicating the removal of the CH$_2$ and CH$_3$ symmetric and asymmetric stretching groups. In addition, the spectra reveal also the disappearance of C-C groups in the wavenumber region between 1350 and 1500 cm$^{-1}$.

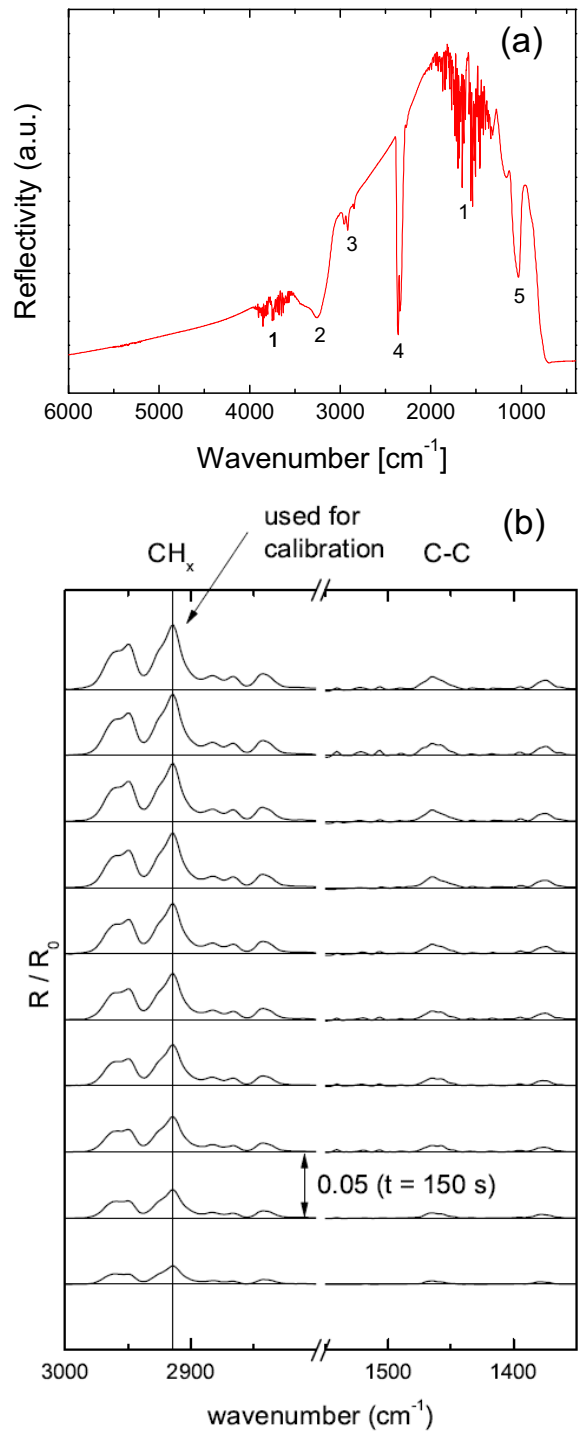

FIG. 7. (a) Reflectance spectrum of a spin-coated PP film on tp of an OCS. The absorption bands are indicated. (b) Temporal evolution of the absorption spectra (from bottom to top) during the exposure of PP to an Ar ion beam with energy of 200 eV.

The reduction of the polymer film thickness during exposure to the particle beams must be implemented in an etch model. Such thickness variation can be calculated from the peak heights by correlating the change in IR signal with the changes in film thickness. This is realized in a calibration experiment by completely etching a film with known thickness, as being determined by AFM. The absolute change in $R/R_0$ of a determined absorption peak can then be correlated to the film thickness. It is worth noting that the gain of IR signal with the OCS method permits to measure a relative change in reflectivity of $\Delta R/R_0 = 10^{-4}$ down to a time resolution of 30 s.

## IV. EXAMPLES

In this section, we describe the application of particle beam experiments to the study of heterogeneous surface processes characteristic of two plasma applications: (i) target oxidation and nitriding in reactive magnetron sputtering, and (ii) plasma treatment of polymers.

### A. Target poisoning during reactive sputtering (Al:O, Al:N)

In reactive magnetron sputtering of Al, the addition of oxygen causes hysteresis effects in the deposition rate due to oxidation of the target surface (target poisoning). One of the consequences is the dramatic reduction in deposition rate. The sputter rate does strongly depend on the surface state of the target: roughness and chemical composition play here a critical role. QCM measurements have been already used to investigate the oxidation processes during target poisoning. Kuschel et al studied the oxidation of Al targets by modeling the surface coverage with oxygen in terms of sputter yields, sticking coefficients and fluxes of the involved particles with Ar ion energies ranging between 200 and 400 eV.[30] There, the same Al electrode of the QCM was used as beam target. The proposed set of balance equations is an extension of the Berg model, which explains the hysteresis effect of the compound target during reactive sputtering.[44] That work introduced in the model an additional term for knock-on implantation of oxygen atoms due to ion bombardment in order to describe the ion-enhanced oxidation of the Al target.

Recently, Kreiter et al. addressed the ion-enhanced oxidation of an Al target by upgrading the previous QCM experiments regarding two aspects: (i) adding an aluminum atom source to mimic also aluminum re-deposition at the target surface during magnetron sputtering,[45] and (ii) investigating the surface with in-situ real time FTIR, to directly monitor the upbuild of oxygen on the surface.[33] The last results were successfully fitted with the same parameters that were used for modeling the mass variation rates measured by QCM. The beam experiments verified thereby the validity of the rate equation model from Kuschel et al. and, moreover, extended it to the poisoned regime of the target. All input parameters such as sputtering yields for Al and oxide as well as sticking coefficients for oxygen (0.015) and aluminum (≈1) are consistent with literature values.

Here we show data measured with the QCM concerning the sputtering of aluminum surface with argon ion and nitrogen as reactive gas. Fig. 8 shows the influence of $N_2$ flux on the sputter yield of an Al target. A smoother dependence of yield with the $N_2$ flux at 600 eV than at 400 eV is observed. Bombardment with nitrogen atoms was achieved by dissociation of $N_2$ molecules using the Evenson cavity. The activation of the microwave cavity resulted in a reduction of ≈70% of the effective sputter yield. Indeed, the bombardment with a nitrogen atom beam may increase the efficiency in surface coverage because of the higher probability of chemisorption of nitrogen atoms on surface sites of the target compared to nitrogen molecules. A rough estimation of N flux in the $N/N_2$ incident beam at the target based on the dissociation degree is made above. Mass spectrometry measurements must be carried out to obtain a real quantification of this collision rate.

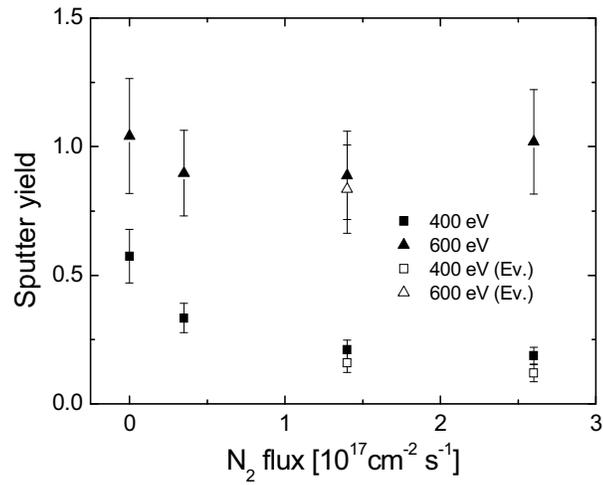

FIG. 8. Influence of the nitrogen flux on the sputter yield of an Al target.

The sticking coefficient of nitrogen atoms on clean Al surfaces was estimated from the data shown in Fig. 9. The measurement started with an Al surface previously cleaned with an $Ar^+$ bombardment until the passivated oxide layer was removed. Afterwards, a constant $N_2$ flux was introduced in the chamber. The gas flow was 3 sccm. The effective sticking coefficient, i.e. the ratio between adsorbed particles and $N_2$ flux, diminishes with time as long as the surface sites are covered with nitrogen atoms. The extrapolation at t=0 provides an approximate value of 0.0002, which is two orders of magnitude lower than the sticking coefficient of oxygen molecules on aluminum (0.015). This very small sticking probability suggests that the deposition of nitrogen atoms is negligible compared to oxidation for similar fluxes of nitrogen and oxygen impinging onto an aluminum surface.

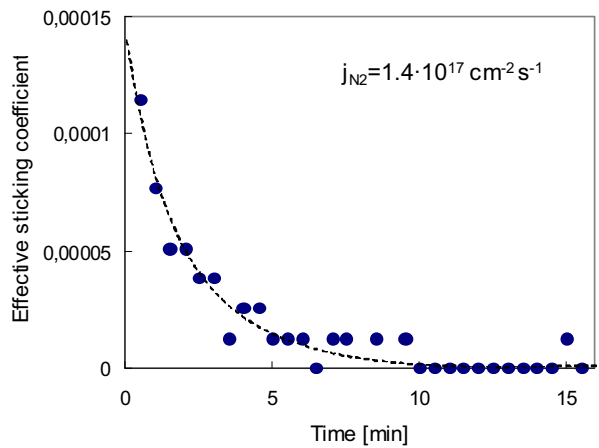

FIG. 9. Evolution of the effective sticking coefficient of $N_2$ on Al exposed to a constant flux of nitrogen. The sticking coefficient at zero surface coverage is determined by extrapolation of the data set as indicated.

## B. Plasma treatment of polymers (PET, PP)

The interaction of plasmas with polymer surfaces is a very common process step in many technologies. Polymers such as PET are coated with silicon oxide layers to enhance their barrier properties or their scratch resistance.[46] In addition, polymers like PP are plasma-treated to make them hydrophilic, as a prerequisite for a subsequent dying, gluing or printing process. In all these processes, modification and etching of an interface on top of the polymer are triggered by ion bombardment in the very first phase of plasma treatment. The effects of ion bombardment must be then described by an adequate model of polymer sputtering. In general, the actual sputtering rate of a given polymer should lie in between two extreme cases: a high yield corresponding to chemical sputtering and a low yield corresponding to physical sputtering.[47-49]

In a previous work, Grosse-Kreul et al have spin-coated PET thin films on an OCS. The samples were exposed to quantified beams of argon ions and of oxygen atoms and molecules.[32] The thickness reduction and change in surface composition were evaluated by real-time by FTIR. The etch rate was modeled with a simple rate equation model based on a balance between the ion-induced cross-linking of the pristine PET film surface, the ion-induced sputtering of this cross-linked surface, and the ion-induced sputtering of the pristine polymer. After the onset of ion bombardment, the etch rate is very high, but decreases rapidly by one order of magnitude to a lower steady state etch rate due to a gradual transition of the pristine polymer into an interconnected surface layer. The etch yields at steady state are of the order of one irrespective of ion energy (range explored: 20-800 eV). Moreover, the addition of oxygen to the incident particle flux is not increasing the etch rate, but only changing the surface composition. The incorporation of oxygen has been confirmed both by a decrease in the C=O absorption group (as determined by FTIR) and by an enhancement in surface energy on a treated sample as measured with the water contact angle method.

By comparison, in this work, PP thin films are exposed to particle beams of argon ions and oxygen at various energies and fluences. Their surface state is continuously monitored by in-situ FTIR. Polymer films of around 30 nm thickness were deposited on OCS by spin-coating, following the procedure described by Song et al.[50] The RMS roughness is of the order of 2-3 nm, as measured by AFM. Analogous to the PET experiments, a transition from a fast initial etch rate to a lower steady state rate was measured when the PP films were bombarded by argon ions. Also, the steady state rates are rather similar, although the ion energy is varied in a broad range, from 20 to 800 eV. The rate equation model of Grosse-Kreul et al was fitted to the experimental data and it revealed sputter yields of the order of 1 for pristine PP at the beginning of the treatment and 0.1 for cross-linked PP in steady state. The contribution of the UV radiation to PP etching was around 0.1 nm/min, very similar to the one measured in PET. The separation of this contribution from the total etch rate provided the sputter yields due only to ions. The values corresponding to cross-linked PP in steady state at low ion energies decreased down to 0.1. In addition, the yields determined from the Monte-Carlo simulation Transport and Range of Ions in Matter (TRIM) are very different from the modeled sputter yields. Such discrepancies owe to the TRIM view of the polymer as an amorphous solid, and the process of chemical sputtering is not taken into account in the simulation.

The influences of atomic and molecular oxygen on PP were investigated using the OBS and the Evenson source. The operation of any of these beam sources did not show any significant modification within the treatment time. As commented above, both sources are expected to deliver similar fluxes of atomic oxygen. Fig. 10 shows, in a first step, how the bombardment of a pristine PP surface only with O atoms (Evenson source) did cause only a mild etching, lower than 0.1 nm/min in the steady state. These rates were calculated from the evolution of the $CH_x$ stretching groups at 2916 and 2954 $cm^{-1}$. As confirmed by in-vacuum XPS measurements using the vacuum carrying case in a later experiment (data not shown here), this observation is interpreted as a net oxygen-induced etching of PP without surface oxidation – no oxygen was detected in the treated sample. In the following step, the treatment with additional $Ar^+$ at ion energy of 20 eV simultaneous to the pure oxygen treatment did not reveal any synergistic effects between oxygen species and VUV photons at the PP surface. In fact, the ion bombardment at such low ion energies is very weak and the ion gun might only act as a VUV source from the ECR plasma.

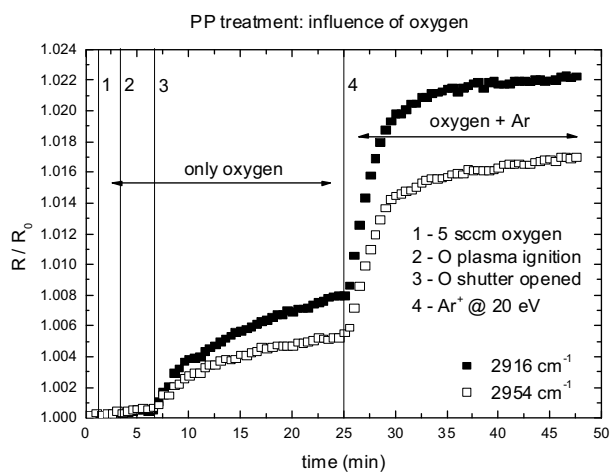

FIG. 10. Variation of the normalized intensities of the groups $CH_x$ during oxygen bombardment with the Evenson source, followed by additional $Ar^+$ treatment from the ion gun.

## V. CONCLUSIONS

Heterogeneous surface processes in plasma applications relevant to reactive sputtering and polymer treatment have been successfully mimicked using a particle beam experiment facility. On one side, QCM measurements have demonstrated that ion-induced oxidation is an important mechanism in target poisoning during reactive magnetron sputtering of aluminum targets. Aluminum poisoning with nitrogen is much weaker compared to oxygen poisoning due to the much reduced sticking coefficient of nitrogen on aluminum surfaces.

On the other hand, irradiation of PET films with argon ions and oxygen atoms and molecules has been monitored by in situ FTIR and shows that chemical sputtering is a dominant process in plasma treatment of this polymer. Surface modification of PP films shows qualitatively similar trends although simultaneous irradiation with argon ions and oxygen atoms does not promote a synergistic effect.

In summary, particle beam experiments with in situ, real time diagnostics by QCM and FTIR are well suited for studying plasma-surface interactions, as the studies of target oxidation and polymer treatment illustrate in this work. Moreover, this kind of experiments can be extended to a variety of research lines due to the flexibility of the reactor set-up.

## ACKNOWLEDGMENTS

This work is supported by the DFG (German Research Foundation) within the framework of the Coordinated Research Centre SFB-TR 87 (project C7) and the Research Department "Plasmas with Complex Interactions" at Ruhr-University Bochum. The authors thank N. Grabkowski and Dr. T. Baloniak for the technical assistance and preparation of some QCM experiments. They are also grateful to Dr. R. Reuter and S. Schröter for the deposition of a-C:H films; to S. Bienholz for the measurements with the RFEA, and to Dr. B.

**REFERENCES**

[1] M. A. Lieberman and A. J. Lichtenberg, *Principles of Plasma Discharges and Materials Processing* (Wiley, New York, 2nd edn. 2005).

[2] Y. Lifshitz, S. R. Kasi and J. W. Rabalais, Phys. Rev. B **41**, 10468 (1990).

[3] R. G. Lacerda, V. Stolojan, D. C. Cox, S. R. P. Silva and F. C. Marques, Diamond Relat. Mater. **11**, 980 (2002).

[4] S. Tajima and K. Komvopoulos, J. Phys. Chem. C **111**, 4358 (2007).

[5] G. S. Oehrlein, R. J. Phaneuf and D. B. Graves, J. Vac. Sci. Technol. B **29**, 010801 (2011).

[6] D. Depla, S. Mahieu and J. E. Greene in P. M. Martin (ed.) *Handbook of Deposition Technologies for Films and Coatings*, 3rd edn. (Elsevier, Amsterdam, 2010), p. 253.

[7] W. Jacob, C. Hopf, A. von Keudell, M. Meier and T. Schwarz-Selinger, Rev. Sci. Instr. **74**, 5123 (2003).

[8] Y. Kimura, J.W. Coburn and D.B. Graves, J. Vac. Sci. Technol. A **22**, 2508 (2004).

[9] H.F. Winters, J.W. Coburn and T.J. Chuang, J. Vac. Sci. Technol. B **1**, 469 (1983).

[10] J.W. Coburn, Pure Appl. Chem. **64**, 709 (1992).

[11] F. Greer, D. Fraser, J.W. Coburn, D.B. Graves, J. Vac. Sci. Technol. A **21** (2003).

[12] S. Chen, H. Kondo, K. Ishikawa, K. Takeda, M. Sekine, H. Kano, S. Den, M. Hori, Jap. J. Appl. Phys. **50**, 01AE03 (2011).

[13] J. P. Allain, M. Nieto, M. R. Hendricks, P. Plotkin, S. S. Harilal and A. Hassanein, Rev. Sci. Instrum. **78**, 113105 (2007).

[14] A. von Keudell, M. Meier and C. Hopf, Diamond Relat. Mater. **11**, 969 (2002).

[15] C. Corbella, S. Portal-Marco, M. Rubio-Roy, E. Bertran, G. Oncins, M.A. Vallvé, J. Ignés-Mullol, J.L. Andújar, J. Phys. D: Appl. Phys. **44**, 395301 (2011).

[16] V.M. Freire, C. Corbella, E. Bertran, S. Portal-Marco, M. Rubio-Roy, J.L. Andújar, J. Appl. Phys. **111**, 124323 (2012).

[17] C. Björkas, K. Vörtler, K. Nordlund, D. Nishijima, R. Doerner, New J. Phys. **11**, 123017 (2009).

[18] D.G. Whyte, G.R. Tynan, R.P. Doerner, J.N. Brooks, Nucl. Fusion **41**, 47 (2001).

[19] M. Missirlian, A. Durocher, A. Grosman, J. Schlosser, J. Boscary, F. Escourbiac, F. Cismondi, Phys. Scr. **T128**, 182 (2007).

[20] A.W. Kleyn, W. Koppers and N. Lopes Cardozo, Vacuum **80**, 1098 (2006).

[21] J. Zhou, I. T. Martin, R. Ayers, E. Adams, D. Liu and E. R. Fisher, Plasma Sources Sci. Technol. **15**, 714 (2006).

[22] I. Tanarro and V. J. Herrero, Plasma Sources Sci. Technol. **18**, 034007 (2009).

[23] D. Gahan, B. Dolinaj and M. B. Hopkins, Rev. Sci. Instrum. **79**, 033502 (2008).

[24] C. Corbella, M. Rubio-Roy, E. Bertran, S. Portal, E. Pascual, M. C. Polo and J. L. Andújar, Plasma Sources Sci. Technol. **20** 015006 (2011).

[25] G. Sauerbrey, Zeit. Phys. **155**, 206 (1959).

[26] K. E. Hurst, A. van der Geest, M. Lusk, E. Mansfield and J. H. Lehman, Carbon **48**, 2521 (2010).

[27] S.S. Narine and A. J. Slavin, J. Vac. Sci. Technol. A **16**, 1857 (1998).

[28] A. Golczewski, A. Kuzucan, K. Schmid, J. Roth, M. Schmid and F. Aumayr, J. Nucl. Mater. **390-391**, 1102 (2009).

[29] V. Raballand, J. Benedikt, J. Wunderlich, A. von Keudell, J. Phys. D: Appl. Phys. **41**, 115207 (2008).

[30] T. Kuschel and A. von Keudell, J. Appl. Phys. **107**, 103302 (2010).

[31] T. Takeuchi, C. Corbella, S. Grosse-Kreul, A. Von Keudell, K. Ishikawa, H. Kondo, K. Takeda, M. Sekine, M. Hori, J. Appl. Phys. **113**, 014306 (2013)

[32] S. Grosse-Kreul, C. Corbella, A. Von Keudell, Plasma Process. Polym. **10**, 225 (2013).

[33] O. Kreiter, S. Grosse-Kreul, C. Corbella, A. Von Keudell, J. Appl. Phys. **113**, 143303 (2013).

[34] T. Baloniak, R. Reuter, C. Flötgen and A. von Keudell, J. Phys. D: Appl. Phys. **43**, 055203 (2010).

[35] http://www.mbe-komponenten.de/products/pdf/data-sheet-obs.pdf

[36] K.G. Tschersich, V. von Bonin, J. Appl. Phys. **84**, 4065 (1998).

[37] F.C. Fehsenfeld, K.M. Evenson, H.P. Broida, Rev. Sci. Instrum. **36**, 294 (1965).

[38] D. Spence, O.J. Steingraber, Rev. Sci. Instrum. **59**, 2464 (1988).

[39] http://www.mbe-komponenten.de/selection-guide/vapor-pressure.php

[40] J. Robertson, Mater. Sci. Eng. R **37**, 129 (2002).

[41] M. A. Hartney, D. W. Hess and D. S. Soane, J. Vac. Sci. Technol. B **7**, 1 (1989).

[42] C. Hopf, M. Schlüter, T. Schwarz-Selinger, U. von Toussaint and W. Jacob, New J. Phys. **10**, 093022 (2008).

[43] A. von Keudell, J.R. Abelson, J. Appl. Phys. **91**, 4840 (2002).

[44] S. Berg, T. Nyberg, Thin Solid Films **476**, 215 (2005).

[45] M. Rossnagel, J. Vac. Sci. Technol. A **6**, 3049 (1988).

[46] M. Deilmann, H. Halfmann, S. Steves, N. Bibinov, P. Awakowicz, Plasma Process. Polym. **6**, S695 (2009).

[47] R.M. France, R.D. Short, S. Robert, H. Building, M. Street, S. Uk, J. Chem. Soc. **93**, 3173 (1997).

[48] J.J. Végh, D. Nest, D.B. Graves, R. Bruce, S. Engelmann, T. Kwon, J. Appl. Phys. **104**, 034308 (2008).

[49] I. Koprinarov, A. Lippitz, J.F. Friedrich, W.E.S. Unger, C. Wöll, Polymer **39**, 3001 (2000).

[50] J. Song, J. Liang, X. Liu, W.E. Krause, J.P. Hinestroza, O.J. Rojas, Thin Solid Films **517**, 4348 (2009).

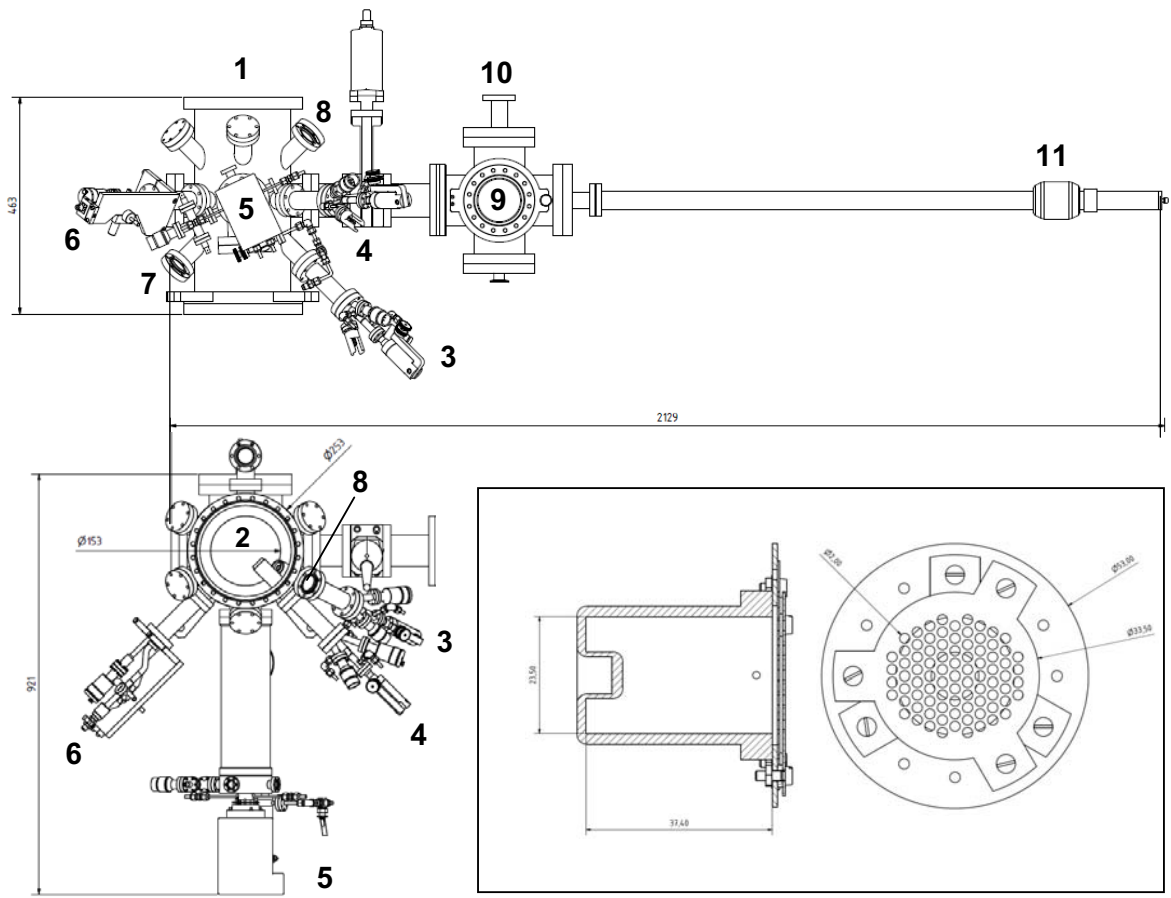

Figure 1 – Corbella et al.

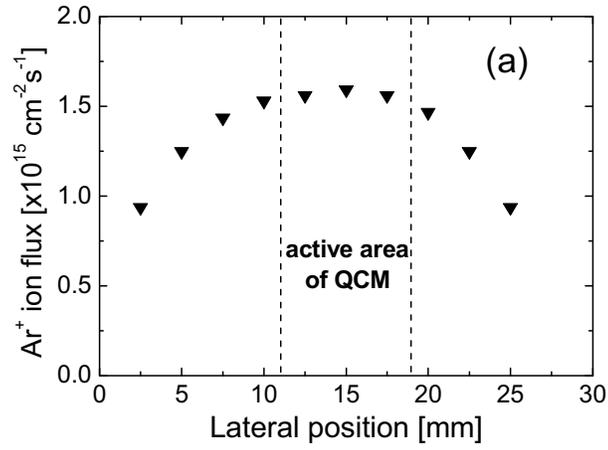
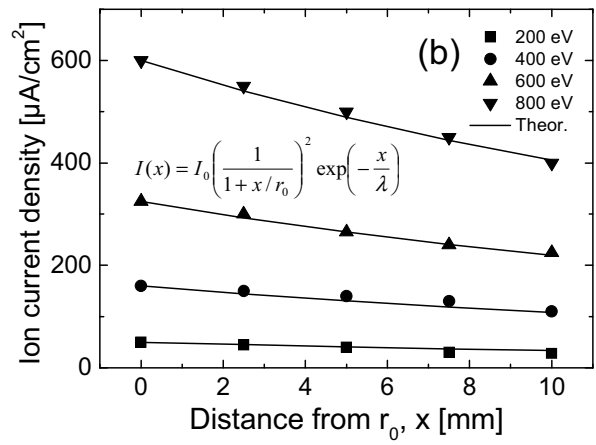

Figure 2 – Corbella et al.

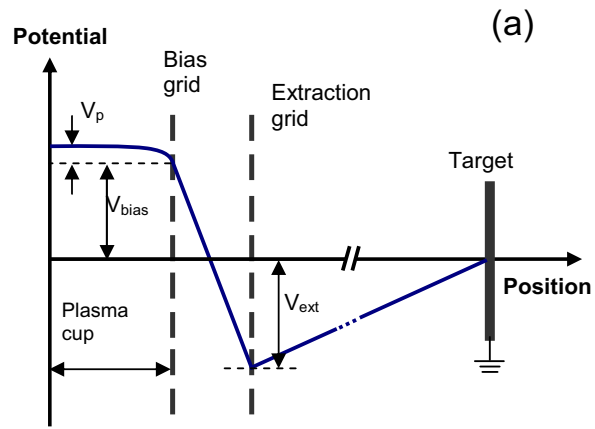
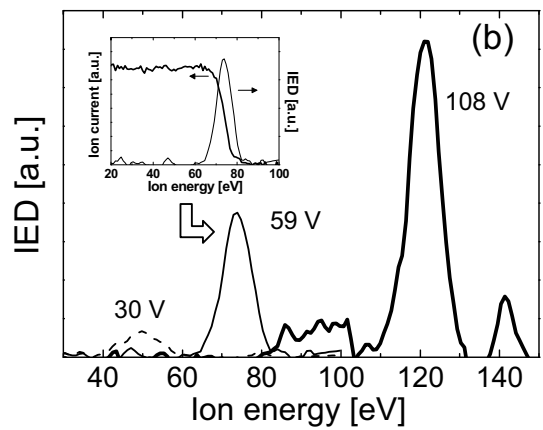

Figure 3 – Corbella et al.

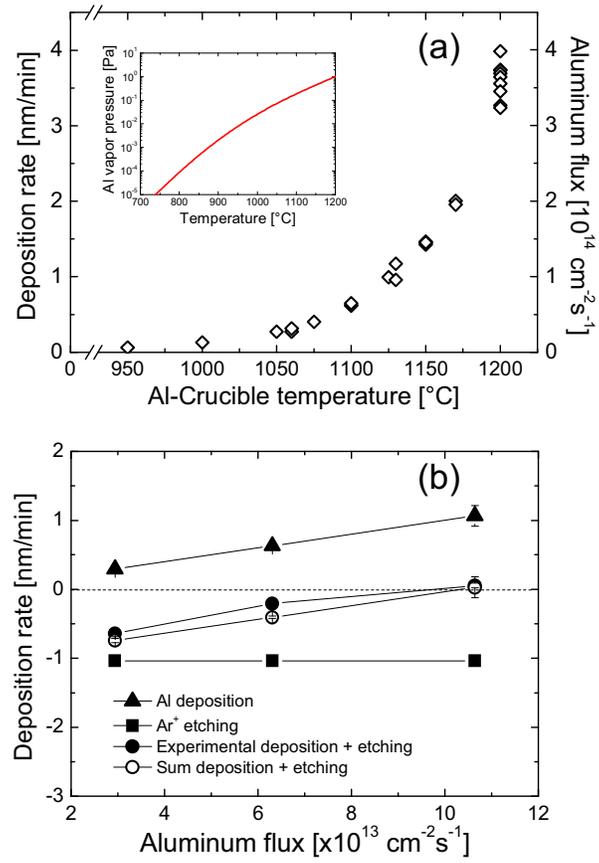

Figure 4 – Corbella et al.

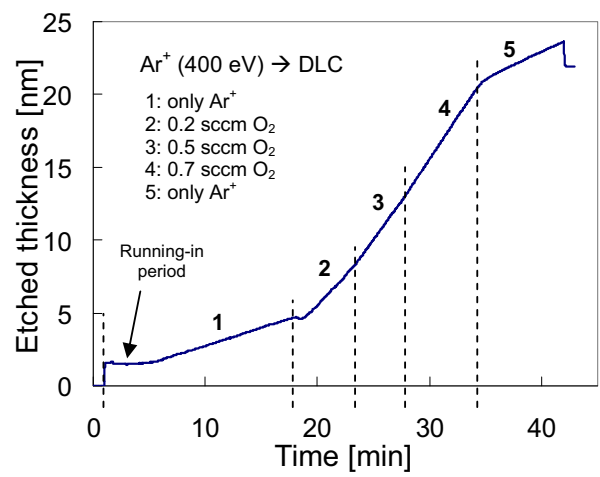

Figure 5 – Corbella et al.

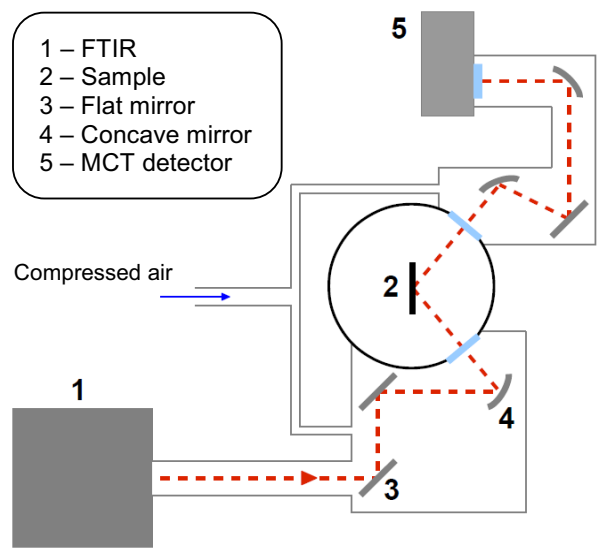

Figure 6 – Corbella et al.

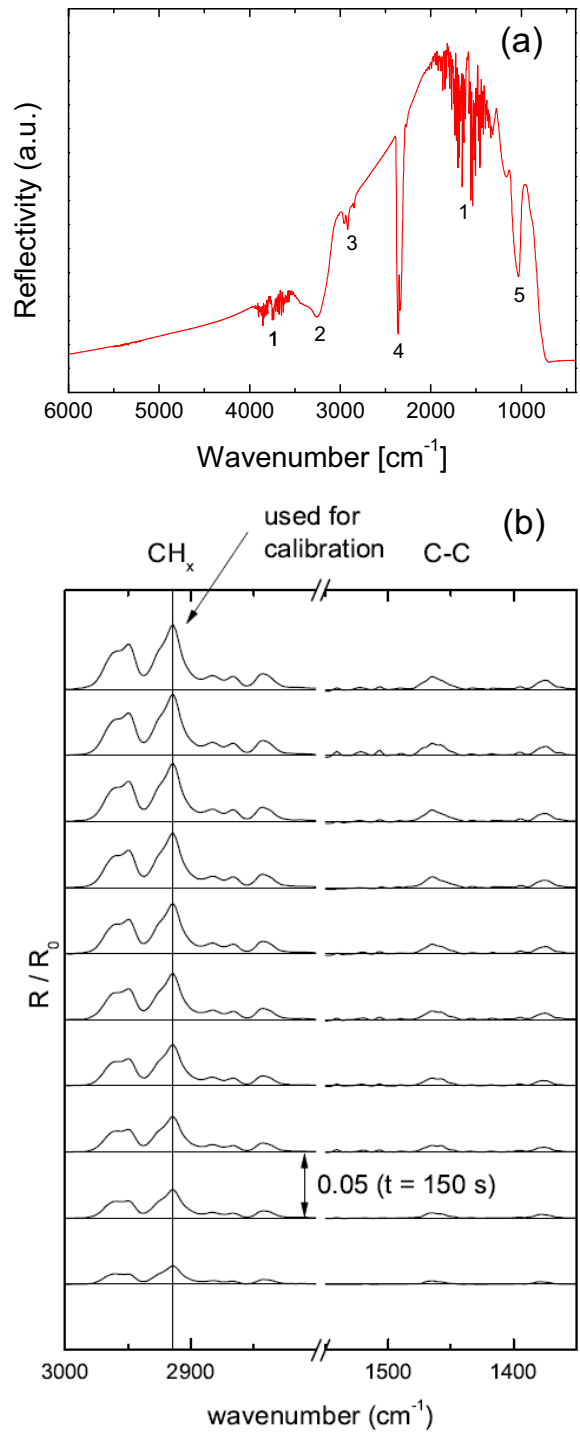

Figure 7 – Corbella et al.

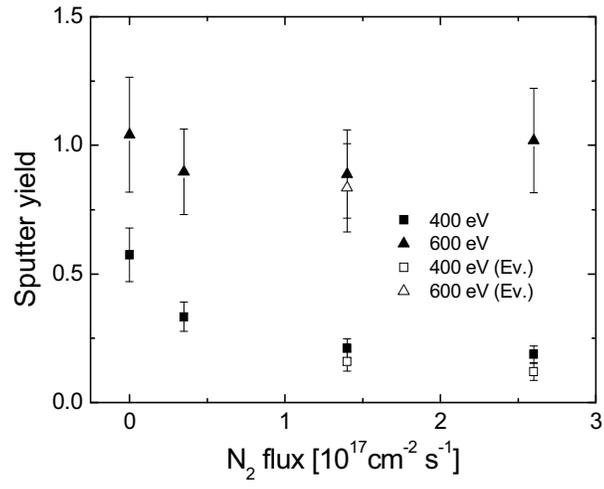

Figure 8 – Corbella et al.

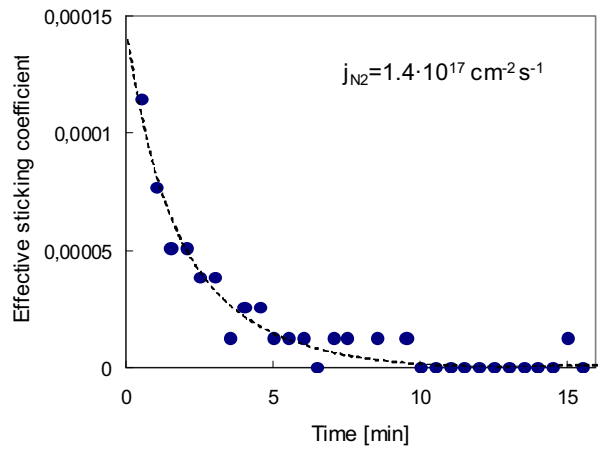

Figure 9 – Corbella et al.

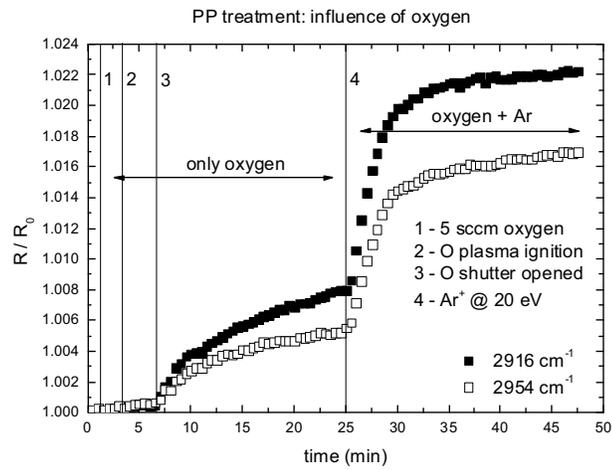

Figure 10 – Corbella et al.